\documentclass[twocolumn, a4paper]{article}
\usepackage{epsfig,amsmath,amsfonts,amssymb,setspace,multirow,textcomp}
\usepackage[T1]{fontenc}
\makeatletter
\renewcommand{\@oddhead}{\textit{Advances in Astronomy and Space Physics} \hfil}
\renewcommand{\@evenfoot}{\hfil \thepage \hfil}
\renewcommand{\@oddfoot}{\hfil \thepage \hfil}
\makeatother

\renewenvironment{thebibliography}[1]{\begin{oldthebibliography}{#1}\setlength{\parskip}{0ex}\setlength{\itemsep}{0ex}}{\end{oldthebibliography}}

\begin{document}
\fontsize{11}{11}\selectfont 
\title{Project VeSElkA: Vertical Stratification of Element Abundances \\ in CP stars\thanks{Based on observations obtained at the Canada-France-Hawaii Telescope (CFHT), which is operated by the National Research Council of Canada, the Institut National des Sciences de l'Univers of the Centre National de la Recherche Scientifique of France, and the University of Hawaii.}}
\author{\textsl{V.\,R.~Khalack$^{1}$, F. LeBlanc$^{1}$}}
\date{\vspace*{-6ex}}
\maketitle
\begin{center} {\small $^{1}$D\'epartement de Physique et d'Astronomie, Universit\'e de Moncton, Moncton, N.-B., Canada E1A 3E9\\
{\tt khalakv@umoncton.ca\\ francis.leblanc@umoncton.ca}}
\end{center}

\begin{abstract}
 A new research project on spectral analysis that aims to characterize the vertical stratification of element abundances in stellar atmospheres of chemically peculiar (CP) stars is discussed in detail.
Some results on detection of vertical abundance stratification in 
several slowly rotating main sequence CP stars are
presented and considered as an indicator of the effectiveness of the atomic diffusion mechanism responsible for the observed peculiarities of chemical abundances. This study is
carried out in the frame of Project VeSElkA (Vertical Stratification of Elements Abundance) for which 34 slowly rotating CP stars have been observed with the ESPaDOnS spectropolarimeter at CFHT.\\[1ex]
\end{abstract}

\section*{\sc introduction}
\indent \indent At certain stages of stellar evolution, some stars show peculiarity of spectral lines in their spectra that argues in favour of an enhanced or depleted abundance of several chemical species in their stellar atmospheres with respect to the solar abundance. Abnormally strong lines of silicon in $\alpha^2$~CVn were first reported by Antonia Maury \cite{Maury1897}, while the strong lines of ionised silicon and ionised strontium were found by Annie Cannon \cite{Cannon1901} in some bright southern stars in the process of determining their spectral class. The first systematic study and classification of stars of spectral classes B-F with abnormally strong lines of various chemical species was performed by Morgan \cite{Morgan1931, Morgan1933}. To explain the mystery of the observed abnormally strong lines, Shapley \cite{Shapley1924} suggested the possibility of abundance abnormalities in the atmospheres of these stars. Several decades later, this idea was confirmed by the results of a differential coarse analysis using the curves of growth \cite{Burbidge1955}. It should be noted, that the observed peculiar abundance of chemical species is only related to the stellar atmosphere and does not reflect the chemical composition of the entire star.

We turn our attention to two types of stars with abundance anomalies: blue horizontal-branch (BHB) stars that burn helium in their core and hydrogen in a shell \cite{Moehler 2004}, and main sequence stars that burn hydrogen in their core. BHB stars are found mostly in globular clusters. Comprehensive surveys show that the hot BHB stars have abundance anomalies as compared to the cool BHB stars in the same cluster \cite{Glaspey+89, Grundahl+99}. Hot BHB stars show enhanced abundances of iron, magnesium, titanium and phosphor, and depleted helium \cite{Behr03a, Behr03b}. Surveys of rotational velocities of BHB stars indicate that stars with $T_{\rm eff} \geq $ 11500~K possess modest rotation with V$\sin{i}<$ 10 km s$^{-1}$, while the cooler stars rotate more rapidly on average \cite{Peterson+95, RB+04}. Slow rotation is an indicator of a hydrodynamically stable atmosphere and the observed abundance peculiarities as well as the brighter $u$-magnitudes (as compared to the theoretical prediction for $u$-magnitudes based on stellar evolution models) of hot BHB stars \cite{Grundahl+98} can be explained in the frame of atomic diffusion mechanism \cite{Michaud1970} at play in their atmospheres. Competition between the gravitational and radiative forces in a hydrodynamically stable atmosphere can cause accumulation or depletion of chemical elements at certain optical depths and lead to a vertical stratification of element abundances. Hui-Bon-Hoa et al. \cite{HBH+00} have constructed stellar atmosphere models of hot BHB stars with vertical stratification of elements, that successfully reproduced some of the anomalies mentioned above. Analysing the available high resolution spectra of some hot BHB stars, Khalack et al. \cite{Khalack+08, Khalack+10} have detected vertical stratification of the abundance of several chemical elements, including iron. The stellar atmosphere models of hot BHB stars calculated self-consistently with abundance stratifications predict that the iron abundance stratification decreases as a function of $T_{\rm eff}$ and becomes negligible for BHB stars hotter than $T_{\rm eff} \simeq $ 14000~K \cite{LeBlanc+10}. This theoretical result is consistent with the detected iron abundance slopes found by Khalack et al. \cite{Khalack+08, Khalack+10} from spectral line analysis of hot BHB stars with different effective temperatures.

A significant portion of upper main sequence stars show abundance peculiarities of various chemical species and are commonly named chemically peculiar (CP) stars \cite{Preston1974}. George Preston \cite{Preston1974} introduced this collective term and divided the CP stars in four groups: CP1 metallic-line stars (type Am-Fm), CP2 magnetic peculiar B and A-type stars (type Bp-Ap), CP3 mercury-manganese stars (type Hg-Mn) and CP4 helium weak stars. Following a large number of extensive studies of CP stars, new types of abundance peculiarities have been discovered and an extension of Preston's classification was proposed \cite{Maitzen1984, Smith1996}. In this new scheme, the CP4 group includes only magnetic He-weak stars (Sr-Ti-Si branch), while the non-magnetic He-weak stars (P-Ga branch) are placed in the group CP5. The He-rich magnetic stars (usually having spectral class B2) are placed in the group CP6. The CP7 group is reserved for the non-magnetic He-rich stars. In this classification the odd numbers correspond to non-magnetic stars, while groups with the even numbers contain stars that possess a detectable magnetic field. More details about the history of discovery and study of CP stars can be found in \cite{Hearnshaw14}.

An extensive list of known and suspected Ap, HgMn and Am stars was recently published by Renson \& Manfroid \cite{Renson+Manfroid09}, while a list of known magnetic CP stars has been compiled by Bychkov et al. \cite{Bychkov+03}. Some upper main sequence CP stars show variability of absorption line profiles in their spectra with the period of stellar rotation due to horizontal inhomogeneous distributions of elements' abundance in their stellar atmosphere \cite{Khokhlova1975}. One can see the periodic (days, months, years) variation of spectral line profiles with the axial rotation of a CP star due to the variation of contribution from different parts of the stellar atmosphere to these lines. Ap stars also show signatures of a strong magnetic field and its structure correlates with the patches of overabundances (or underabundances) of certain elements \cite{Khochukhov02, Shavrina+10}. Observationally, it was shown that the properties of the line profile variability due to abundance patches as well as the magnetic field structure of Ap stars do not change during several decades  \cite{Mathys+Hubrig97, Romanyuk+14} and, therefore, it is usually assumed that stellar atmospheres of Ap stars are hydrodynamically stable. In a hydrodynamically stable atmosphere the presence of a magnetic field can intensify accumulation or depletion of chemical elements at certain optical depths \cite{Alecian+Stift10, Ryabchikova+08, Stift+Alecian12}. From the analysis of line profiles of ionised iron and ionised chromium in the spectra of $\beta$~CrB, Ryabchikova et al. \cite{Ryabchikova+03} have shown that iron and chromium abundances increase towards the deeper atmospheric layers. This can be explained in terms of the mechanism of atomic diffusion \cite{Michaud1970}. It appears that the light and the iron-peak elements are concentrated in the lower atmospheric layers of some Ap stars (HD~133792 and HD~204411) \cite{Ryabchikova+04}. Meanwhile the rare-earth elements (for example, Pr and Nd) are usually pushed into the upper atmosphere of Ap stars \cite{Mashonkina+05}.

Detection of vertical stratification of element abundances in stellar atmospheres of stars is an indicator of the effectiveness of the atomic diffusion mechanism that may be responsible for the observed abundance peculiarities. The vertical stratification of element abundances in a hydrodynamically stable atmosphere can be estimated through the analysis of multiple line profiles that belong to the same ion of the studied element \cite{Khalack+Wade06, Khalack+07} using the ZEEMAN2 code \cite{Landstreet88}, for instance. This approach has been successfully employed to study the vertical abundance stratification in the atmospheres of several BHB \cite{Khalack+08, Khalack+10} and HgMn stars \cite{Thiam+10}.


To search for the signatures and to study the abundance stratification of chemical species with optical depth in the atmospheres of CP stars, we have initiated Project VeSElkA, which means rainbow in ukrainian and stands for Vertical Stratification of Element Abundances. Slowly rotating (V$\sin{i}$ < 40 km/s) CP stars of the upper main sequence were selected for our study using the catalogue of Ap, HgMn and Am stars of \cite{Renson+Manfroid09}. This limitation for the rotational velocity is imposed with the aim to increase the probability for a star to have a hydrodynamically stable atmosphere, where the atomic diffusion mechanism can produce vertical stratification of element abundances. A low value of V$\sin{i}$ also leads to narrow and mostly unblended line profiles in the observed spectra, which is beneficial for our abundance analysis.




\begin{table}[!th]
 \centering
 \caption{CP stars observed with ESPaDOnS in the frame of Project VeSElkA.}
 \vspace*{1ex}
 \begin{tabular}{lrrl} 
  \hline
  name & $M_{V}$ & V$\sin{i}$, & type \\ 
       &         & km/s &      \\        
  \hline
HD2628   & 5.22 & 21 & A7    \\
HD6397*  & 5.65 & 10 & F3m Sr\\
HD12869* & 5.03 & 18 & A2m   \\
HD15385  & 6.19 & 29 & A5 \\
HD22920  & 5.53 & 39 & B8p Si\\
HD23878  & 5.24 & 24 & A1 \\
HD24712  & 6.00 & 22 & A9p Sr-Cr-Eu\\
HD25267  & 4.66 & 28 & B8.5p Si \\
HD40394  & 5.71 & 18 & B9.5 Si-Fe\\
HD53929  & 6.09 & 25 & B9 Mn \\
HD68351* & 5.61 & 33 & A0 Si-Cr\\
HD71030  & 6.10 &  9 & F6 \\
HD83373  & 6.39 & 28 & A1 Sr-Cr-Eu\\
HD90277  & 4.73 & 34 & F0 \\
HD95608  & 4.41 & 21 & A1m \\
HD97633  & 3.35 & 23 & A2 Sr-Eu \\
HD110380 & 3.48 & 23 & F2m \\
HD116235 & 5.89 & 26 & A2m \\
HD148330 & 5.75 & 18 & A2 Si-Sr \\
HD157087 & 5.37 & 15 & A4 \\
HD158261*& 5.94 & 17 & A1 \\
HD159082*& 6.45 & 22 & B9 HgMn \\
HD164584 & 5.34 & 12 & F3 \\
HD166473 & 7.92 & 20 & A8p Sr-Cr-Eu \\
HD170973 & 6.41 & 18 & A0p Si-Cr-Sr \\
HD174933 & 5.40 & 20 & B9 HgMn \\
HD176232 & 5.89 & 18 & A8 Sr \\
HD186568 & 6.06 & 15 & B9 \\
HD190229*& 5.66 &  8 & B9 HgMn \\
HD191110 & 6.18 &  0 & B9.5 \\
HD196821 & 6.08 & 22 & A0 \\
HD207840 & 5.77 & 15 & B8 Si \\
HD209459 & 5.82 & 14 & B9.5 \\
HD214994 & 4.79 & 14 & A1 \\
HD223640 & 5.18 & 28 & B7p Si-Cr-Sr \\
HD224103 & 6.21 & 28 & B9 Si \\
 \hline
 \end{tabular}
\\ $^{*}$Known spectroscopic binary stars
\label{tab1}
\end{table}

\section*{\sc observations and data reduction}
\label{observ}
\indent \indent The selected slowly rotating CP stars and two normal main sequence stars (used as reference stars) have been observed during 2013-14 with ESPaDOnS
(Echelle SpectroPolarimetric Device for Observations of Stars) at the CFHT (Canada-France-Hawaii Telescope) employing the deep-depletion e2v device Olapa.
The instrument performances as well as the optical characteristics of the spectropolarimetre are described in \cite{Donati+06}.\footnote{
For more details about this instrument, the reader is invited to visit {\rm www.cfht.hawaii.edu/Instruments/Spectroscopy/Espadons/}}

ESPaDOnS acquires high resolution (R=~65000) Stokes IV spectra throughout the spectral range from 3700\AA\, to 10500\AA\, in a single exposure \cite{Donati+99}.
In order to be able to find convincing signatures of vertical stratification of element abundances from the spectral analysis, we require spectra with high signal-to-noise ratio (close to one thousand per bin in the spectral region around 5150\AA). Table~\ref{tab1} presents a list of the observed slowly rotating chemically peculiar stars and some normal upper main sequence stars. The first, second and third columns provide respectively the name of a star, its apparent visual magnitude and V$\sin{i}$ value, while the fourth column contains information about its spectral type.  
The spectral classes of stars selected for the Project VeSElkA are given in Table~\ref{tab1} taking into account the data from \cite{Renson+Manfroid09}, SIMBAD Astronomical Database\footnote{http://simbad.u-strasbg.fr/simbad/} and the estimates of effective temperature and surface gravity obtained by Khalack \& LeBlanc \cite{Khalack+LeBlanc15}. Some of the observed program stars are spectroscopic binaries. Our sample also includes several HgMn stars and magnetic Bp-Ap stars. For each star, we have obtained at least two spectra to verify if it is a binary star, or if the observed line profiles are variable with the phase of stellar rotation due to horizontal stratification of element abundances.

The obtained spectra have been reduced using the dedicated software package Libre-ESpRIT \cite{Donati+97} which yields both the Stokes I spectrum and the Stokes V circular polarisation spectrum. The Stokes V spectra are required to estimate and study the stellar magnetic field and its potential contribution to the atomic diffusion mechanism \cite{Stift+Alecian12}. The level of continuum (free from absorbtion and emission lines) in each echelle order of the Stokes I spectra was approximated with a polynomial function and its coefficients were derived. Using this polynomial function, the Stokes IV spectra were normalized resulting in a continuum with variations no larger than two to three percent.
To estimate the effective temperature and gravity of the observed stars (see next section), we have used their non-normalized spectra \cite{Khalack+LeBlanc15}.

\begin{figure}[!t]
\centering
\begin{minipage}[t]{1.0\linewidth}
\centering
\epsfig{file = 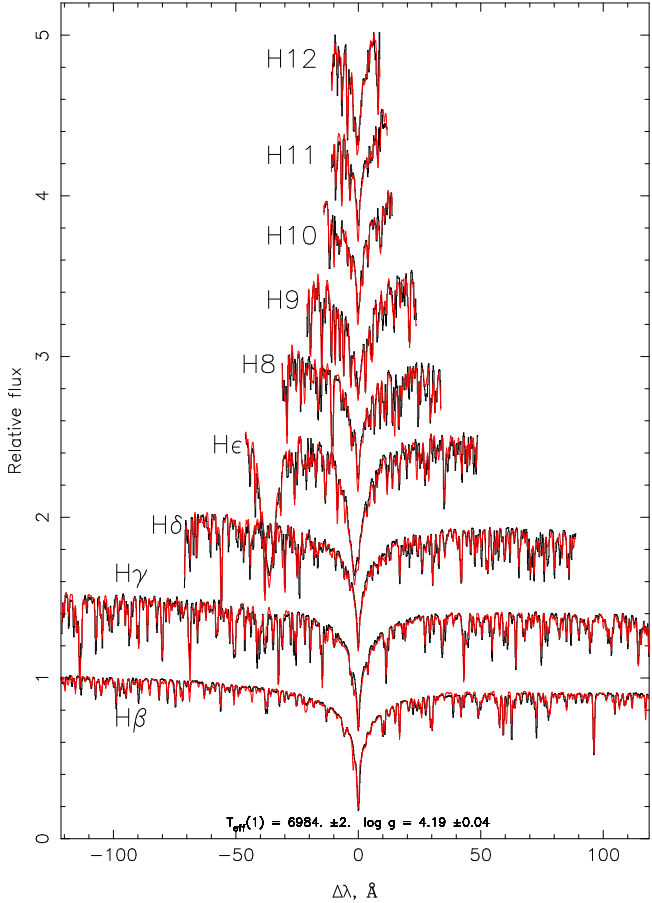,width = 1.0\linewidth}
\caption{The observed spectrum (thick line) of HD~110380 is well fitted by synthetic spectrum (thin dotted line) that corresponds to $T_{\rm eff}$ = 6980$\pm$200 ~K, $\log{g}$ = 4.19$\pm$0.2 ($\chi^2/\nu$ = 0.31) \cite{Khalack+LeBlanc15}.}\label{fig1}
\end{minipage}
\end{figure}

\section*{\sc estimation of fundamental parameters of stellar atmospheres }
\label{param}
\indent \indent The abundance analysis of normalized polarimetric spectra has been carried out employing a modified version \cite{Khalack+Wade06, Khalack+07} of the ZEEMAN2 radiative transfer code \cite{Landstreet88}, which requires atomic data and a synthetic stellar atmosphere model (local temperature, pressure, electronic density, etc. relative to atmospheric optical depth) characterized by the effective temperature, surface gravity and metallicity to simulate the synthetic line profiles.
The atomic data provided by the VALD-2 \cite{Kupka+99} and NIST \cite{Kramida+13} databases have been used for our simulations.
To determine the parameters of stellar atmospheres, one can fit the observed Balmer line profiles with synthetic spectra from a grid of stellar atmosphere models with different values of $T_{\rm eff}$, $\log{g}$ and metallicity. The stellar atmosphere models used here were calculated with the PHOENIX atmospheric code \cite{Hauschildt+97}.

With the aim to estimate the effective temperature, surface gravity and metallicity of stars observed in the frame of the Project VeSElkA, we have generated a new library of high resolution (0.05\AA\, in the visible range from 3700\AA\, to 7700\AA) synthetic spectra \cite{Khalack+LeBlanc15} using version fifteen of the PHOENIX code \cite{Hauschildt+97}. A grid of stellar atmosphere models and corresponding fluxes has been calculated for 5000~K $\leqslant T_{\rm eff} <$ 9000~K with a step of 250~K, for 9000~K $\leqslant T_{\rm eff} \leqslant$ 15000~K with a step of 500~K and for 3.0 $\leqslant \log{g} \leqslant$ 4.5 with a step of 0.5. The grids of models have been produced for the solar metallicity \cite{Grevesse+10} as well as for the metallicities [M/H]= -1.0, -0.5, +0.5, +1.0, +1.5 assuming a nil microturbulent velocity.

Nine Balmer line profiles observed in the non-normalized spectra of several CP stars selected for Project VeSElkA were fitted using the code FITSB2 \cite{Napiwotzki+04} to determine the fundamental parameters of their stellar atmosphere. An example of the best fit of Balmer lines in HD~110380 obtained for the solar metallicity is presented in Fig.~\ref{fig1}. This is a F2m star (see Table.~\ref{tab1}) with $T_{\rm eff}$ = 6980 K, $\log{g}$ = 4.19 \cite{Khalack+LeBlanc15}, for which our best fit approximates quite well also the Ca\,{\sc ii} 3934\AA\, and 3968\AA\, line profiles in the left side of the Balmer $H_{\epsilon}$ line and in its left wing respectively (see Fig.~\ref{fig1}).

We have found that our grids provide almost the same values of fundamental parameters of stellar atmospheres for the studied CP stars as do the Atlas9 grids calculated by Castelli \& Kurucz \cite{C+K04}. However, while studying the sensitivity of the determined values of $T_{\rm eff}$ and $\log{g}$ to the set of Balmer lines used, we have found that the use of the Atlas9 grids may produce some ambiguity in the determination of fundamental stellar parameters if the effective temperature is close to 10000~K depending on which set of Balmer lines is used. Meanwhile, the Phoenix-15 grids are not sensitive to the choice of Balmer lines in the range of effective temperatures from 9700~K to 12000~K \cite{Khalack+LeBlanc15}.

\begin{figure}[!t]
\centering
\begin{minipage}[t]{1.0\linewidth}
\centering
\epsfig{file = 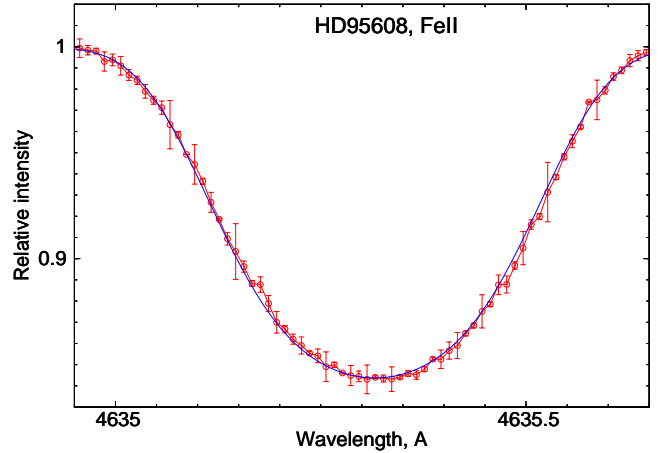,width = 0.70\linewidth,angle=-90}
\caption{Example of the best fit (line) obtained for the combined Fe\,{\sc ii} 4635\AA\, line profile (open circles) in HD~95608 \cite{Khalack+14}. The vertical error-bars represent the measurement precision of normalised flux.}
\label{fig2}
\end{minipage}
\end{figure}

\section*{\sc abundance analysis}
\label{analysis}
\indent \indent From the analysis of observed line profiles while using the modified ZEEMAN2 code \cite{Khalack+Wade06, Khalack+07}, we can determine the radial velocity of the studied star, the projection of its rotational velocity to the line of sight V$\sin{i}$ and the abundance of the chemical element responsible for this absorption line. Fig~\ref{fig2} shows an example of the Fe\,{\sc ii} 4635\AA\, line profile observed in HD~95608 that is well fitted by the theoretical profile \cite{Khalack+14}. 
Preliminary analysis has shown that iron lines are not variable in the HD~95608 spectra obtained in a time span of few weeks. Therefore, in the case of HD~95608, we can use the line profile composed from the data of several spectral observations for our analysis (see Fig~\ref{fig2}).

For the abundance analysis, we usually select unblended (not contaminated by a contribution from other chemical species) absorbtion line profiles that belong to a particular ion. In our study, we assume that the core of the line profile is formed mainly at line optical depth $\tau_{\rm \ell}$=1, which corresponds to
a particular layer of the stellar atmosphere.
Therefore, from the simulation of each line profile that belongs to a particular ion, we can obtain its abundance at a particular layer of the stellar atmosphere (that corresponds to a particular continuum optical depth $\tau_{\rm 5000}$).

Taking into account that the analysed lines usually have different lower energetic levels and oscillator strengths, their cores are generally formed at different optical depths $\tau_{\rm 5000}$. In this way, we can study the vertical distribution of an element's abundance from the analysis of a large number (at least ten or more) line profiles that belong to one or two ions of this element that are detected in the analyzed spectrum \cite{Khalack+08}. A statistically significant vertical stratification of the element's abundance is considered when the abundance change with optical depth exceeds 0.5 dex. More details on the fitting procedure are given in \cite{Khalack+07}. This method was successfully used by Khalack et al. \cite{Khalack+08, Khalack+10} to study the vertical abundance stratification of chemical species in BHB stars and by Thiam et al. \cite{Thiam+10} in HgMn stars.

\begin{figure}[!t]
\centering
\begin{minipage}[t]{1.0\linewidth}
\centering
\epsfig{file = 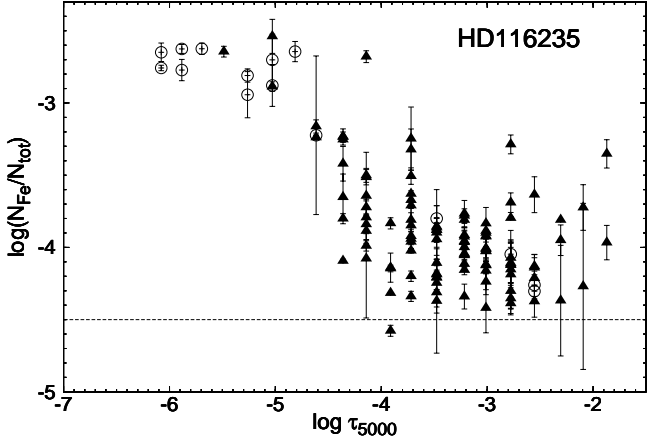,width = 0.70\linewidth,angle=-90}
\caption{The iron vertical stratification detected from the analysis of individual lines observed in the spectra of HD~116235. The filled triangles represent lines of neutral iron while the open circles stand for singly ionized iron \cite{Khalack+14}. The dashed line indicates the iron abundance in the solar atmosphere \cite{Grevesse+10}.}\label{fig3}
\end{minipage}
\end{figure}

\section*{\sc results from Project VeSElkA}
\label{results}
\indent \indent Using our grids of stellar atmosphere models and synthetic spectra calculated with the PHOENIX-15 code \cite{Hauschildt+97} we have estimated the values of effective temperature, surface gravity and metallicity for sixteen stars observed in the frame of VeSElkA Project \cite{Khalack+LeBlanc15}. Our results on $T_{\rm eff}$ and $\log{g}$  obtained for twelve of these stars are consistent with the previously published data. Meanwhile, for the four other stars (HD~23878, HD~83373, HD~95608 and HD~164584) we have for the first time reported the estimates of their effective temperature, gravity and metallicity in \cite{Khalack+LeBlanc15}.

Estimation of average abundances and detection of vertical abundance stratification of some chemical species were carried out for several CP stars selected for Project VeSElkA \cite{Khalack+14, LeBlanc+15}.
Clear evidences of vertical stratification of iron and chromium abundances were found in the stellar atmospheres of HD~95608 and HD~116235 (see Fig.~\ref{fig3} for an example).
For the analysed chemical species, no evidence of vertical stratification was found in the atmospheres of HD~71030 and HD~186568 \cite{Khalack+14, LeBlanc+15}.
The radial and rotational velocities determined for the four stars under consideration 
are consistent with the values found in previous studies. According to LeBlanc et al.~\cite{LeBlanc+15}, HD~71030
does not show large abundance anomalies and it might be a normal main sequence star.

Two normal B-type stars HD~186568 and HD~209459 \cite{Dworetsky+Budaj00} have been selected for Project VeSElkA as the reference stars (see Table~\ref{tab1}). We plan to use them to test the applied method for the detection of vertical abundance stratification and to verify our estimates of average abundance of chemical species well represented by a large number of spectral lines in the analysed spectra. As was mentioned above, we have not found signatures of vertical abundance stratification from the analysis of HD~186568 spectra. The solar abundances of titanium and iron were also obtained for HD~209459, which is known to have no abundance peculiarities \cite{Hubrig+Castelli01}.

\begin{figure}[!t]
\centering
\begin{minipage}[t]{1.0\linewidth}
\centering
\epsfig{file = 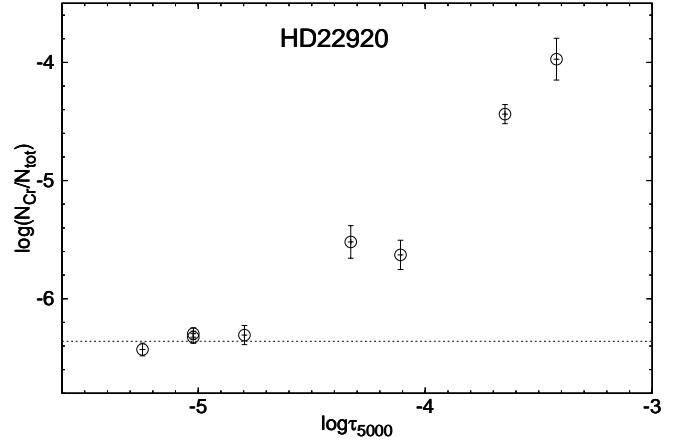,width = 0.70\linewidth,angle=-90}
\caption{The increase of chromium abundance towards the deeper atmospheric layers in HD~22920. Each open circle represents the data obtained from the analysis of distinct line profile that belongs to Cr\,{\sc ii} ion \cite{Khalack+Poitras15}. The dashed line indicates the chromium abundance in the solar atmosphere \cite{Grevesse+10}.}\label{fig4}
\end{minipage}
\end{figure}


The average abundances of oxygen, silicon, iron and chromium were obtained for different rotational phases of HD~22920 that shows clear signatures of the presence of a moderate magnetic field. All the analyzed elements show variability of their line profiles with the rotational phase. This argues in favour of a non-uniform horizontal distribution of their abundances. Among the studied elements, the Si\,{\sc ii} line profiles show the strongest variability with rotational phase \cite{Khalack+Poitras15}.

To study the vertical abundance stratification of chemical species, the observed line profiles were fitted with the synthetic ones, simulated assuming a homogeneous horizontal distribution of element abundances and no magnetic field. This simplification is not physically correct, but for a given rotational phase it provides an estimate for the average abundances of chemical species. Meanwhile, the contribution of the magnetic field is accounted for by the slightly larger value of V$\sin{i}$ obtained as the final result of fitting.
Among the chemical species represented by a number of unblended spectral lines, only silicon and chromium appear to show strong signatures of vertical stratification of their abundance in the stellar atmosphere of HD~22920 \cite{Khalack+Poitras15}. Fig.~\ref{fig4} shows that chromium has a tendency to increase its abundance towards the deeper atmospheric layers in HD~22920. Similar behaviour with atmospheric depth was also found for the silicon abundance \cite{Khalack+Poitras15}.

\section*{\sc discussion}
\label{discuss}
\indent \indent Accumulation or depletion of chemical elements at certain optical depths brought about by atomic diffusion \cite{Michaud1970} can modify the structure of stellar atmospheres of CP stars \cite{LeBlanc+09, Stift+Alecian12} and lead to the observed horizontal \cite{Khochukhov02, Shavrina+10} and vertical \cite{Ryabchikova+04, Mashonkina+05} stratification of element abundances. For the Bp-Ap stars, it is believed that the substantial magnetic field suppresses convection and contribute significantly to the mechanism of atomic diffusion \cite{Stift+Alecian12}. In this way, the magnetic field can stabilise the structure of abundance peculiarities in atmospheres of magnetic CP stars, so that they remain stable over more than fifty years of spectral observations \cite{Mathys+Hubrig97, Romanyuk+14}. Therefore, for a comprehensive study of the structure of stellar atmospheres in CP stars, it is important to detect and observationally estimate the intensity of vertical abundance stratification of different chemical species \cite{Khalack+LeBlanc15, LeBlanc+15}. Such detections are the main goal of Project VeSElkA. These results may lead to improved calculations of the self-consistent stellar atmosphere models with vertical stratification of elements using the PHOENIX code \cite{LeBlanc+09}, or other atmospheric models such as those of Stift \& Alecian \cite{Stift+Alecian12}.

For several program CP stars, we have estimated their effective temperature, surface gravity and metallicity through the fitting of nine Balmer line profiles with synthetic spectra calculated for a grid of stellar atmosphere models. For this aim, a new library of grids of stellar atmosphere models and corresponding fluxes has been generated and tested \cite{Khalack+LeBlanc15}. The obtained stellar atmosphere parameters have been used to calculate homogeneous stellar atmosphere models, that were employed to perform an abundance analysis of HD~22920, HD~23878, HD~95608, HD~116235 and HD~186568 \cite{Khalack+14, Khalack+Poitras15, LeBlanc+15} with the help of the modified ZEEMAN2 code \cite{Khalack+Wade06, Khalack+07}. Among the studied CP stars, signatures of vertical stratification of silicon and chromium abundances were found in HD~22920 \cite{Khalack+Poitras15}, and of iron and chromium abundances in HD~95608 and HD~116235 \cite{Khalack+14, LeBlanc+15}.

\begin{figure}[!t]
\centering
\begin{minipage}[t]{1.0\linewidth}
\centering
\epsfig{file = 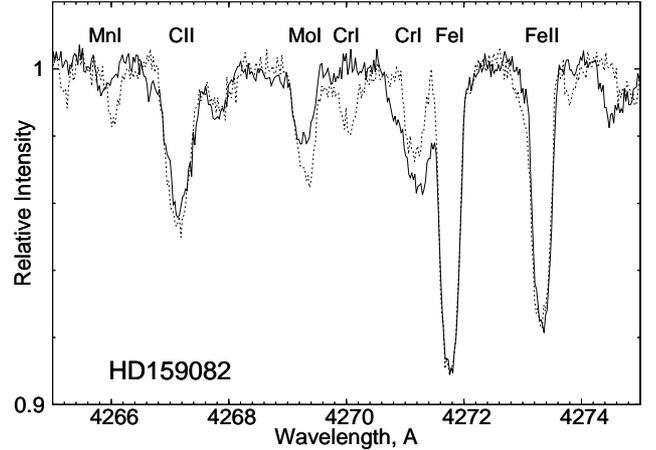,width = 0.70\linewidth,angle=-90}
\caption{Example of overlapping of two spectra corrected for radial velocity of the main component in the spectroscopic binary HD~159082.}\label{fig5}
\end{minipage}
\end{figure}

In order to improve the precision of measurement of the abundance stratification with atmospheric depths for different chemical elements we need to analyse more unblended spectral line profiles that belong to each element. Therefore, we plan to expand our study towards the near infrared spectral region, where absorbtion lines of many metals can be found for the cool upper main sequence stars (spectral classes A0-F5). For this aim, we look forward to analyse the spectra from the near IR spectropolarimetre/velocimetre SPIRou \cite{Barrick+12} proposed to be installed in 2017 at the CFHT. The Canada Foundation for Innovation (CFI) has approved the production of a copy of SPIRou that is planned to be installed at a telescope with high aperture located in the southern hemisphere.
The spectropolarimetre SPIRou will allow the acquisition of high resolution (R>70000) Stokes IVQU spectra throughout the spectral range from 0.98 to 2.5 $\mu$m (YJHK bands) in a single exposure \cite{Delfosse+13}. Such high spectral resolution in the near IR will provide the measurement of radial velocities with a precision higher than one meter per second \cite{Delfosse+13}. Taking into account the aforementioned characteristics of SPIRou, we consider it to be very useful to search for signatures of vertical abundance stratification in stars. The high resolution near IR spectra of CP stars, even with the relatively low signal-to-noise ratio, in combination with their optical spectra could significantly improve the precision of the statistical evaluation of vertical stratification of element abundances.

The abundance analysis of other CP stars selected for Project VeSElkA is underway. The use of available high-resolution and high signal-to-noise spectra of CP stars obtained by other astronomers will be analysed 
to accumulate an extensive database for vertical stratification of element abundances in these stars. The database will be used to search for a potential dependence of the vertical abundance stratification of different chemical species relative to the effective temperature similar to the one that we have found for BHB stars \cite{LeBlanc+10}.

We are also going to study the spectra of some magnetic Bp-Ap stars with the aim to detect vertical stratification of element abundances and to search for possible correlation of the abundance peculiarities with the magnetic field structure. Special attention will be paid to the CP stars in binary systems, because due to the tidal interaction its members usually have a slow axial rotation \cite{Abt+02, Song+13}, which can lead to a hydrodynamically stable atmosphere. Fig.~\ref{fig5} shows an example of two overlapping spectra corrected for radial velocity of the main component in the spectroscopic binary HD~159082. One can see there, that the Fe\,{\sc i} and Fe\,{\sc ii} line profiles are relatively stable and can be used for the analysis of vertical abundance stratification. Meanwhile, the Cr\,{\sc i}, Mo\,{\sc i}, Ni\,{\sc i} and He\,{\sc i} line profiles change significantly during a period of time shorter than one day. The spectra of the binary stars found in Table~\ref{tab1} have been incorporated in the database of the BinaMIcS project \cite{Alecian+15} that provides high quality spectra of magnetic CP stars in binary systems.

\section*{\sc acknowledgement}
\indent \indent We are thankful to the Facult\'{e} des \'{E}tudes Sup\'{e}rieures et de la Recherche de l'Universit\'{e} de Moncton and NSERC for research grants. The grids of stellar atmosphere models and corresponding fluxes have been generated using the supercomputer {\it briarree} of l'Universit\'{e} de Montr\'{e}al, that operates under the guidance of Calcul Qu\'{e}bec and Calcul Canada. The use of this supercomputer is funded by the Canadian Foundation for Innovation (CFI), NanoQu\'{e}bec, RMGA and Research Fund of Qu\'{e}bec - Nature and Technology (FRQNT). This paper has been typeset from a \TeX/\LaTeX\, file prepared by the authors.



\begin{thebibliography}{3}
{\small


\bibitem{Abt+02} Abt~H.\,A., Levato~H., Grosso~M. 2002, ApJ, 573, 359
\bibitem{Alecian+Stift10} Alecian~G., \& Stift~M.\,J. 2010, A\&A, 516, 53
\bibitem{Alecian+15} Alecian~E., Neiner~C., Wade~G.\,A. et al. 2015, in '{\it New windows on massive stars: asteroseismology, interferometry and spectropolarimetry}', Proc. of IAUS 307, eds.: G.~Meynet, C.~Georgy, J.\,H.~Groh \& Ph.~Stee, 330
\bibitem{Barrick+12} Barrick~G.\,A., Vermeulen~T., Baratchart~S., et al. 2012, in '{\it Software and Cyberinfrastructure for Astronomy II.}', Proc. of SPIE, 8451, 15, [article id. 84513J]
\bibitem{Behr03a} Behr~B.\,B. 2003a, ApJS, 149, 67
\bibitem{Behr03b} Behr~B.\,B. 2003b, ApJS, 149, 101
\bibitem{Burbidge1955} Burbidge~G.\,R. \& Burbidge~E.\,M. 1955, Astrophys. J. Suppl. Ser., 1, 431
\bibitem{Bychkov+03} Bychkov~V.\,D., Bychkova~L.\,V., \& Madej~J. 2003, A\&A 407, 631
\bibitem{Cannon1901} Cannon~A.\,J. \& Pickering~E.\,C.  1901, Annals of the Astronomical Observatory of Harvard College, 28, 129
\bibitem{C+K04} Castelli~F., \& Kurucz~R.\,L., 2003, in '{\it Modelling of stellar atmospheres}', Proc. of IAUS 210, eds.: N.~Piskunov, W.\,W.~Weiss, \& D.\,F.~Gray, A20
\bibitem{Delfosse+13} Delfosse~X., Donati~J.-F., Kouach~D. et al. 2013, in Proc. of the Annual meeting of the French Soc. of Astronomy and Astrophysics, SF2A-2013, eds.:
L.~Cambresy, F.~Martins, E.~Nuss, \& A.~Palacios, 497
\bibitem{Donati+97} Donati~J.-F., Semel~M., Carter~B.\,D., et al., 1997, MNRAS, 291, 658
\bibitem{Donati+99} Donati~J.-F., Catala~C., Wade~G.\,A., et al., 1999, A\&ASS, 134, 149
\bibitem{Donati+06} Donati~J.-F., Catala~C., Landstreet~J.\,D., Petit~P., \& ESPaDOnS team 2006, in '{\it Solar Polarization 4}', ASP Conf. Series, 358, eds.: R.~Casini, \& B.\,W.~Lites, 362
\bibitem{Dworetsky+Budaj00} Dworetsky~M.\,M.,\&  Budaj~J., 2000, MNRAS, 318, 1264
\bibitem{Glaspey+89} Glaspey~J.\,W., Michaud~G., Moffat~A.\,F.\,J., \& Demers~S. 1989, ApJ, 339, 926
\bibitem{Grevesse+10} Grevesse~N., Asplund~M., Suaval~A.\,J., \& Scott~P. 2010, Ap\&SS, 328, 179
\bibitem{Grundahl+98} Grundahl~F., Vandenberg~D.\,A, Andersen~M.\,I. 1998, ApJ, 500, L179
\bibitem{Grundahl+99} Grundahl~F., Catelan~M., Landsman~W.\,B., Stetson~P.\,B., \& Andersen~M. 1999, ApJ, 524, 242
\bibitem{Hauschildt+97} Hauschildt~P.\,H., Shore~S.\,N., Schwarz~G.\,J., et al., 1997, ApJ, 490, 803
\bibitem{Hearnshaw14} Hearnshaw~J.\,B. 2014, '{\it The analysis of starlight: two centuries of astronomical spectroscopy}', Cambrige University Press, New York
\bibitem{Hubrig+Castelli01} Hubrig~S., \& Castelli~F., 2001, A\&A, 375, 963
\bibitem{HBH+00} Hui-Bon-Hoa~A., LeBlanc~F., Hauschildt~P.\,H. 2000, ApJ, 535, L43
\bibitem{Khalack+Wade06} Khalack~V., \& Wade~G. 2006, A\&A, 450, 1157
\bibitem{Khalack+07} Khalack~V., LeBlanc~F., Bohlender~D., Wade~G.\,A., Behr~B.\,B. 2007, A\&A, 466, 667
\bibitem{Khalack+08} Khalack~V.\,R., LeBlanc~F., Behr~B.\,B., Wade~G.\,A., Bohlender~D. 2008, A\&A, 477, 641
\bibitem{Khalack+10} Khalack~V.\,R., LeBlanc~F., Behr~B.\,B. 2010, MNRAS, 407, 1767
\bibitem{Khalack+14} Khalack~V., Yameogo~B., Thibeault~C., LeBlanc~F. 2014, in '{\it Magnetic fields throughout stellar evolution}', Proc. of IAUS 302, eds.: P.~Petit, M.~Jardine, \& H.\,C.~Spruit, 272
\bibitem{Khalack+Poitras15} Khalack~V., \& Poitras~P. 2015, in '{\it New windows on massive stars: asteroseismology, interferometry and spectropolarimetry}', Proc. of IAUS 307, eds.: G.~Meynet, C.~Georgy, J.\,H.~Groh \& Ph.~Stee, 383
\bibitem{Khalack+LeBlanc15} Khalack~V.\,R., LeBlanc~F. 2015, AJ, accepted, [arXiv:1505.08158]
\bibitem{Khochukhov02} Khochukhov~O., Piskunov~N., Ilyin~I., Ilyina~S., Tuominen~I. 2002, A\&A, 389, 420
\bibitem{Khokhlova1975} Khokhlova~V.\,L. 1975, Astron. Zh., 52, 950
\bibitem{Kramida+13} Kramida~A., Ralchenko~Yu., Reader~J., \& NIST ASD Team 2013, NIST Atomic Spectra Database (ver. 5.1). Available: http://physics.nist.gov/asd. National Institute of Standards and Technology, Gaithersburg, MD
\bibitem{Kupka+99} Kupka~F., Piskunov~N., Ryabchikova~T.\,A., Stempels~H.\,C., Weiss~W.\,W., 1999, A\&AS, 138, 119
\bibitem{Landstreet88} Landstreet~J.\,D. 1988, ApJ, 326, 967
\bibitem{LeBlanc+09} LeBlanc~F., Monin~D., Hui-Bon-Hoa~A., Hauschildt~P.\,H., 2009, A\&A, 495, 937
\bibitem{LeBlanc+10} LeBlanc~F., Hui-Bon-Hoa~A., Khalack~V. 2010, MNRAS, 409, 1606
\bibitem{LeBlanc+15} LeBlanc~F., Khalack~V., Yameogo~B., Thibeault~C., Gallant~I. 2015, MNRAS, submitted
\bibitem{Maitzen1984} Maitzen~H.\,M. 1984, A\&A, 138, 493
\bibitem{Mashonkina+05} Mashonkina~L., Ryabchikova~T., Ryabtsev~A. 2005, A\&A, 441, 309
\bibitem{Mathys+Hubrig97} Mathys~G., Hubrig~S. 1997, A\&AS, 124, 475
\bibitem{Maury1897} Maury~A\,C. \& Pickering~E.\,C.  1897, Annals of the Astronomical Observatory of Harvard College, 28, 1
\bibitem{Michaud1970} Michaud~G. 1970, ApJ, 160, 641
\bibitem{Moehler 2004} Moehler~S., 2004, in '{\it The A-star puzzle}', Proc. of IAUS 224, eds.: Zverko~J., Weiss~W.\,W., \v Zi\v z\v novsk\'y~J., \& Adelman~S.\,J., 119
\bibitem{Morgan1931} Morgan~W.\,W. 1931, Astrophys. J., 73, 104
\bibitem{Morgan1933} Morgan~W.\,W. 1933, Astrophys. J., 77, 77
\bibitem{Napiwotzki+04} Napiwotzki~R., Yungelson~L., Nelemans~G., et al. 2004, in '{\it Spectroscopically and Spatially Resolving the Components of the Close Binary Stars}', ASP Conf. Series, 318, eds.: R.\,W.~Hilditch, H.~Hensberge \& K.~Pavlovski, 402
\bibitem{Peterson+95} Peterson~R.\,C., Rood~R.\,T., \& Crocker~D.\,A. 1995, ApJ, 453, 214
\bibitem{Preston1974} Preston~G.\,W. 1974, Astron. Astrophys. Suppl. Ser. 12, 257
\bibitem{Romanyuk+14} Romanyuk~I.\,I., Semenko~E.\,A., Kudryavtsev~D.\,O. 2014, Astrophysical Bulletin, 69, 4, 427
\bibitem{Ryabchikova+03} Ryabchikova~T., Wade~G.\,A.; LeBlanc~F. 2003, in '{\it Modelling of stellar atmospheres}', Proc. of IAUS 210, eds.: N.~Piskunov, W.\,W.~Weiss, \& D.\,F.~Gray, 301
\bibitem{Ryabchikova+04} Ryabchikova~T., Leone~F., Kochukhov~O., Bagnulo~S. 2004, in '{\it The A-star puzzle}', Proc. of IAUS 224, eds.: Zverko~J., Weiss~W.\,W., \v Zi\v z\v novsk\'y~J., \& Adelman~S.\,J., 580
\bibitem{Ryabchikova+08} Ryabchikova~T., Kochukhov~O., Bagnulo~S. 2008, A\&A, 480, 811
\bibitem{RB+04} Recio-Blanco~A., Piotto~G., Aparicio~A.,\& Renzini~A. 2004, A\&A, 417, 597
\bibitem{Renson+Manfroid09} Renson~P., Manfroid~J. 2009, A\&A 498, 961
\bibitem{Shapley1924} Shapley~H. 1924, Harvard Coll. Observ. Bull. 798, 2
\bibitem{Shavrina+10} Shavrina~A.\,V., Glagolevskij~Yu.\,V., Silvester~J., et al. 2010, MNRAS, 401, 1882
\bibitem{Smith1996} Smith~K.\,C. 1996, Asptrophys. \& Space Sci., 237, 77
\bibitem{Song+13} Song~H.\,F., Maeder~A., Meynet~G., et al. 2013, A\&A, 556, 100
\bibitem{Stift+Alecian12} Stift~M.\,J., Alecian~G. 2012, MNRAS, 425, 2715
\bibitem{Thiam+10} Thiam~M., LeBlanc~F., Khalack~V., Wade~G.\,A. 2010, MNRAS, 405, 1384
}
\end{thebibliography}
\end{document}